\documentstyle{elsart}

\newcommand{\beq}[1]{\begin{equation} \label{#1} }
\newcommand{\eeq}   {\end{equation}}
\newcommand{\Frac}[2]{\frac{\textstyle \mathstrut #1}
{\textstyle \mathstrut #2}}
\newcommand{\ve}{\varepsilon}
\newcommand{\vkappa}{\mbox{\boldmath $\kappa$}}
\newcommand{\vrho}{\mbox{\boldmath $\rho$}}
\newcommand{\av}[1]{\langle #1 \rangle}
\begin{document}

\begin{frontmatter}

\title{Interaction of Ultra-Cold Neutrons with Condensed Matter}
\author{S.T.Belyaev} and \author{A.L.Barabanov}
\address{RRC "Kurchatov Institute", 123182 Moscow, Russia}
\date{}

\begin{abstract}
General theory of neutron scattering (elastic and inelastic) is
presented. It is applicable for the whole domain of slow neutrons
and includes as limiting cases existing theories for thermal and
cold neutrons and for elastic scattering of UCN. New expression
for inelastic scattering cross section for UCN is proposed. It
differs from the usually used by proper account of re-scattering
processes. Evidence for small heating and cooling of UCN is given.
\end{abstract}

\end{frontmatter}

PACS number(s): 25.40.Fq, 61.12.Bt


\section{Introduction}

Thermal and cold neutrons with wave length
$0.03$~nm~$\leq\lambda\leq1$~nm is an important tool for
investigation of condensed matter. Theory of their interaction
with substance is well established (see, e.g., \cite{1,2}). It is
based on the use of Fermi pseudopotential. For thermal and cold
neutrons re-scattering of secondary waves is unimportant and one
may use Born approximation that gives for double differential
cross-section the following expression
\beq{1}
\frac{d^2\sigma}{d\Omega d\omega}=
\frac{k'}{2\pi k}\sum_{\nu\nu'}
b_{\nu}^*b_{\nu'}\chi_{\nu\nu'}(\vkappa,\omega).
\eeq
Here $\vkappa={\bf k}'-{\bf k}$, $\omega=\ve-\ve'$, where
${\bf k}$ and $\ve$ are momentum and energy of incident neutron,
${\bf k}'$ and $\ve'$ are the same quantities for scattered
neutron, and $b_{\nu}$ is a scattering amplitude on bound
$\nu$-th nucleus. A Fourier transform
\beq{2}
\chi_{\nu\nu'}(\vkappa,\omega)=
\int\limits_{-\infty}^{+\infty}
\chi_{\nu\nu'}(\vkappa,t)e^{i\omega t}dt
\eeq
of diagonal matrix element of an operator of nuclear position
correlation
\beq{3}
\chi_{\nu\nu'}(\vkappa,t)=
\av{i|e^{i\vkappa\hat {\bf R}_{\nu}(t)}
e^{-i\vkappa\hat {\bf R}_{\nu'}(0)}|i}
\eeq
between the initial eigenfunctions $|i\rangle$ of target
Hamiltonian, that determines the target response on
the scattered neutron wave.

For ultra-cold neutrons (UCN), when $\lambda\ge 10$~nm,
re-scattering of neutron wave in media is very essential, and when
$k^2<4\pi bn$ re-scattering becomes the dominant process and
results in the total reflection from the surface of the target
(of cause, for positive $b$). Thus, Born approximation in general,
and the cross-section (\ref{1}) in particular, can not be used for
UCN. To describe an elastic scattering of UCN by matter one uses a
multiple scattering wave approach for fixed (unmovable) nuclei
(see, e.g., \cite{3,4}). It gives an effective repulsive (optical)
potential for neutron inside a condensed matter, so the neutron wave
with the energy below the threshold is exponentially decreasing
deep into target. However, UCN escape from vessels, that attracts
attention of experimenters for many years, as well as small
heating and cooling observed recently \cite{5}, belong to
inelastic processes. In this paper we present the basic features
of a general theory for elastic and inelastic scattering equally
applicable for thermal and cold as well as ultra-cold neutrons,
and which, when neutron wave-length decreases, smoothly transforms
into the usual scattering theory giving
the cross-section~(\ref{1}).

\section{ General expressions}

A proper theory for UCN scattering should be based on the
following postulates: (i) No Born approximation; (ii) No use of
Fermi potential; (iii) Target matter is a dynamical system. It is,
of cause, impossible to solve the many-body problem of
neutron -- target interaction without any approximations. In our
problem there are two main small parameters: short-range of
neutron-nuclei interaction (as compared with interatomic distance
and wave length), and small neutron energy (as compared with depth
of interaction potential).

The first condition allows to consider only s-wave part of the
wave function of neutron -- nucleus center-of-mass motion, when
their interaction is evaluated. And the second condition allows in
this evaluation to neglect energy of relative neutron -- nucleus
motion inside the interaction potential area. No specific model
for neutron -- nucleus interaction potential is needed. Its
specific features described above (short range and large depth)
allows to use scattering length approximation.

From these considerations we obtain a general expression for
double differential cross section
\beq{4}
\frac{d^2\sigma}{d\Omega d\omega}=
\frac{k'}{2\pi k}\sum_{jj'\nu\nu'}
\phi^{j*}_{\nu}\phi^{j'}_{\nu'}
\chi^{jj'}_{\nu\nu'}(\vkappa,\omega+E_i-E_j).
\eeq
It contains neutron amplitudes $\phi^j_{\nu}$ and Fourier
transform of nondiagonal matrix element of
the correlation operator
\beq{5}
\chi^{jj'}_{\nu\nu'}(\vkappa,t)=
\av{j|e^{i\vkappa\hat {\bf R}_{\nu}(t)}
e^{-i\vkappa\hat {\bf R}_{\nu'}(0)}|j'}
\eeq
between the eigenfunctions $|j\rangle$ and $|j'\rangle$ of target
Hamiltonian. Note, that $E_i$ is the energy of the initial target
state $|i\rangle$, and $E_j$ corresponds to a state $|j\rangle$. A
set of linear algebraic equations for neutron amplitudes
$\phi^j_{\nu}$ is also found.

Neglecting in these equations terms that describe
re-scattering we get for the amplitudes
\beq{6}
\phi^j_{\nu}=\delta_{ij}\,\beta_{\nu}
\left(1-i\alpha_{\nu}\av{i|\sqrt{\hat{\bf k}_{\nu}^2}\,|i}\right),
\eeq
where $\alpha_{\nu}$ and $\beta_{\nu}$ are scattering lengths on
isolated and bound nucleus, respectively, and $\hat{\bf k}_{\nu}$
is an operator of impact momentum in the center-of-mass system for
neutron and nucleus. Thus, for thermal and cold neutrons
Eq.(\ref{4}) really transforms into Eq.(\ref{1}), and the usual
relation between $\beta_{\nu}$ and $b_{\nu}$ arises.

Into a condensed matter we have
${\bf R}_{\nu }=\vrho_{\nu }+{\bf u}_{\nu }$, where $\vrho_{\nu }$
is the equilibrium position of the $\nu$-th nucleus, and
${\bf u}_{\nu }$ is its shift from the equilibrium. Thus, the
factors $e^{-i{\bf k}\vrho_{\nu}}$ and $e^{i{\bf k}\vrho_{\nu'}}$
may be extracted in the matrix element (\ref{5}) and combined with
the amplitudes $\phi^{j*}_{\nu}$ and $\phi^{j'}_{\nu'}$ in
(\ref{4}). Equations for new amplitudes
$\psi^j(\nu)=(\phi^j_{\nu}/\beta_{\nu})e^{i{\bf k}\vrho_{\nu}}$
are of the form
\beq{7}
\psi^j(\nu)=\delta_{ij}e^{i{\bf k}\vrho_{\nu}}-
\sum_{j'\nu'}\beta_{\nu'}G^{jj'}_{\nu\nu'}\psi^{j'}(\nu').
\eeq
The coefficients $G^{jj'}_{\nu\nu'}$ are expressed in terms of
matrix elements (\ref{5}).

Then we use an expansion over ${\bf k}{\bf u}$ for functions
$\chi^{jj'}_{\nu\nu'}(\vkappa,\omega)$, coefficients
$G^{jj'}_{\nu\nu'}$, and amplitudes $\psi^j(\nu)$ (or
$\phi^j_{\nu}$). Zero-order approximation (${\bf u}=0$)
corresponds to fixed nuclei and, therefore, results in only
elastic scattering. Equations for zero-order amplitudes
$\psi^{(0)j}(\nu)=\delta_{ij}\psi_{{\bf k}}(\nu)$
\beq{8}
\psi_{{\bf k}}(\nu)=e^{i{\bf k}\vrho_{\nu}}-
\sum_{\nu'}\beta_{\nu'}G^i(\nu\nu')\psi_{{\bf k}}(\nu'),\qquad
G^i(\nu\nu')=\frac{e^{ik|\vrho_{\nu}-\vrho_{\nu'}|}}
{|\vrho_{\nu}-\vrho_{\nu'}|}.
\eeq
coincide with the multiple scattering wave equations usually used
to describe UCN elastic scattering.

\section{Inelastic scattering}

Inelastic scattering arises in the second-order
approximation in ${\bf k}{\bf u}$.
Analysis shows that there are four second-order terms
in inelastic cross section~(\ref{4})
\beq{9}
\begin{array}{l}
\sum\limits_{jj'}\phi^{j*}_{\nu}\phi^{j'}_{\nu'}
\chi^{jj'}_{\nu\nu'}\quad\to\quad
\phi^{(0)i*}_{\nu}\phi^{(0)i}_{\nu'}
\chi^{(2)ii}_{\nu\nu'}+
\phi^{(0)i*}_{\nu}\phi^{(1)f}_{\nu'}
\chi^{(1)if}_{\nu\nu'}+{}
\\
\phantom{\phi^{j*}_{\nu}\phi^{j'}_{\nu'}
\chi^{jj'}_{\nu\nu'}\quad\quad}+
\phi^{(1)f*}_{\nu}\phi^{(0)i}_{\nu'}
\chi^{(1)fi}_{\nu\nu'}+
\phi^{(1)f*}_{\nu}\phi^{(1)f}_{\nu'}
\chi^{(0)ff}_{\nu\nu'}.
\end{array}
\eeq
Four terms in the right-hand side of (\ref{9}) are illustrated
by Fig.1.
To disclose physical meaning of these terms, it is instructive
to compare our result with that based on the improvement of
(\ref{1}) by replacement of Born amplitudes $b_{\nu}$ by
the neutron amplitudes in optical potential $\phi^{(0)}_{\nu}$
(see, e.g., \cite{6}). In such an approach expansion similar
to (\ref{9}) would evidently result in only the first term
(Fig.1a), where re-scattering is taken into account only for
incident neutron wave (already included in $\phi^{(0)}_{\nu}$).

Three other terms in (\ref{9}) describe rescattering of
out-going waves (in inelastic channels). They are directly
and indirectly generated by nondiagonal matrix element 
$\chi^{jj'}_{\nu\nu'}$. The first-order term for diagonal
matrix element ($j=j'$) is absent.

Final expression for the second order inelastic cross section is
of the form
\beq{10}
\frac{d\sigma^{(2)}_{ie}}{d\omega}=
\frac{1}{2\pi mk}\int d^3k'\,\delta(\ve'+\omega-\ve)
\int\frac{d^3q}{(2\pi)^3}B^*_{\alpha}({\bf q})B_{\beta}({\bf q})
\Omega_{\alpha\beta}({\bf q},\omega),
\eeq
\beq{11}
{\bf B}({\bf q})=
\sum_{\nu}\beta_{\nu}e^{-i{\bf q}\vrho_{\nu}}
\nabla_{\nu}\left(\psi_{-{\bf k}'}(\nu)\psi_{{\bf k}}(\nu)\right),
\eeq
where $\Omega_{\alpha\beta}({\bf q},\omega)$ is related with
Fourier transform of correlation function by the equation
\beq{12}
\av{i|\hat u_{\nu\alpha}(t)\hat u_{\nu'\beta}(0)|i}=
\int\frac{d^3qd\omega}{(2\pi)^4}
e^{i{\bf q}(\vrho_{\nu}-\vrho_{\nu'})-i\omega t}
\Omega_{\alpha\beta}({\bf q},\omega).
\eeq
Note, that (\ref{11}) contains symmetrically the functions of the
elastic and inelastic neutron channels. In the Born approximation,
i.e., neglecting by re-scattering both in the elastic and
inelastic
channels, we have from (\ref{8}):
$\psi_{{\bf k}}(\nu)\to e^{i{\bf k}{\vrho_{\nu}}}$ and
$\psi_{-{\bf k}'}(\nu)\to e^{-i{\bf k}'{\vrho_{\nu}}}$. Thus,
\beq{13}
{\bf B}({\bf q})\enskip\to\enskip -i\vkappa
\sum_{\nu}\beta_{\nu}e^{-i({\bf q}+\vkappa)\vrho_{\nu}},
\eeq
and the usually used formula for inelastic cross section arises
(see, e.g., \cite{4}). In \cite{6} an attempt was made to improve
this approach by replacing the plane wave
$e^{i{\bf k}{\vrho_{\nu}}}$ in (\ref{13}) by the damping function
$\psi_{{\bf k}}(\nu)$. This attempt is clearly inconsistent as
such replacement should be made in (\ref{11}) before
differentiation with respect to $\vrho_{\nu}$.

\section{Results}

To illustrate the possibilities of our approach we studied small
heating and cooling of UCN in a simplest model. Let us consider
UCN that fall normally to a thick layer of uniform matter with an
energy $\ve$ below a threshold $U$. Taking the correlation
function in phonon model we obtain for the probabilities of
inelastic scattering per one bounce the following expressions
\beq{14}
\left.\frac{dw^{(2)}_{ie}}{d\ve'}\right|_{\ve'\le U}=
\frac{2k\beta}{\pi U}\,\frac{T}{Ms^2}\,\frac{v'}{s},\quad
\left.\frac{dw^{(2)}_{ie}}{d\ve'}\right|_{\ve'>U}=
\frac{k\beta}{\pi \ve'}\,\frac{T}{Ms^2}\,
\frac{\Frac{v'}{s}}{1-\left(\Frac{v'}{2s}\right)^2},
\eeq
where $T$ is a target temperature, $M$ is a mass of target nuclei,
$s$ is a speed of sound, and $v'=\sqrt{2\ve'/m}$ is a velocity of
a scattered neutron. Note, that the second formula is not valid in
a small region just above the barrier, where oscillations governed
by narrow resonances in transmission and reflection are of
importance. However, these oscillations damp rapidly when $\ve'$
increases. A half of neutrons with the energy $\ve'>U$ reflects
from the target, while the other half transmits through it.

Spectrum of inelastically scattered neutrons in the model
considered has a maximum at $\ve'\simeq U$. Indeed, it increases
as $\sim v'$ below the barrier and falls off as $\sim 1/v'$ above
the barrier. Thus, qualitatively it is just of the form needed to
explain small heating and cooling of UCN into vessels. However,
the magnitude of the effect in the phonon model is low, as
$k\beta\sim 10^{-6}$ and $v'/s\sim 10^{-3}$. Nevertheless, it
should be noted that an evaluation of the inelastic  scattering
probability in the Born approximation, i.e., with
${\bf B}({\bf q})$ (\ref{13}), gives
\beq{15}
\frac{dw^{B}_{ie}}{d\ve'}\sim
\frac{k\beta}{ms^2}\,\frac{T}{Ms^2}\,\frac{v'}{s}.
\eeq
This spectrum, first, has no maximum in low energy region and,
second, is additionally suppressed by the factor $\sim U/ms^2$ as
compared with Eqs.(\ref{14}). This results from the direct
proportionality of ${\bf B}({\bf q})$ (\ref{13}) to $\vkappa$
vanishing for low energy transfer from neutron to target or vice
versa.

One should expect that the low energy transfer processes are
governed not by phonons but rather by other collective excitations
in condensed matter. In particular, when a propagation speed of
the excitation is of the same order as the velocity of UCN, an
influence of matter fluctuations on re-scattering processes may be
maximal. Our study of UCN interaction with diffusion and thermal
wave modes is now in progress.

\section{Summary}

General theory of neutron scattering (elastic and inelastic) is
presented. It is applicable for the whole domain of slow neutrons
and includes as limiting cases existing theories for thermal and
cold neutrons and for elastic scattering of UCN. The only small
parameters used are those for the interaction potential that was
assumed a short and relatively deep, what is equivalent to
scattering length approximation for the interaction. Evident
expression for the inelastic cross section is given. It differs
from the usually used by proper account of re-scattering in the
inelastic channel. It is shown that in the phonon model our
approach qualitatively explains the low energy transfer processes.
However, to provide the large observed probabilities of small
heating and cooling of UCN into vessels other collective
excitations of condensed matter in the limit of small ${\bf q}$
and $\omega$ should be apparently taken into account.

\begin{ack}
This work was supported by RFBR Grant 96-15-96548.
\end{ack}

\newpage

\centerline{FIGURE CAPTURE}
\bigskip

Fig.1. Contributions to the second-order inelastic cross section:
(a) scattering -- scattering interference,
(b) scattering -- re-scattering interference,
(c) re-scattering -- re-scattering interference.
Solid and dash lines represent neutrons in the elastic and
inelastic channels, respectively. Open and crossed circles
correspond to elastic and inelastic scattering, respectively.

\newpage

\begin{center}
\begin{picture}(85,35)
\put(20,15){\circle{5.5}}
\put(18,17){\line(1,-1){4}}
\put(18,13){\line(1,1){4}}
\put(0,35){\vector(1,-1){17.5}}
\multiput(22.5,17.5)(1,1){17}{\circle*{0.5}}
\put(37,32){\vector(1,1){3}}
\put(19,5){$\nu$}
\put(9,15){$|i\rangle$}
\put(65,15){\circle{5.5}}
\put(63,17){\line(1,-1){4}}
\put(63,13){\line(1,1){4}}
\put(45,35){\vector(1,-1){17.5}}
\multiput(67.5,17.5)(1,1){17}{\circle*{0.5}}
\put(82,32){\vector(1,1){3}}
\put(64,5){$\nu'$}
\put(54,15){$|i\rangle$}
\put(42,3){a}
\end{picture}
\end{center}
\bigskip

\begin{center}
\begin{picture}(85,35)
\put(20,15){\circle{5.5}}
\put(18,17){\line(1,-1){4}}
\put(18,13){\line(1,1){4}}
\put(0,35){\vector(1,-1){17.5}}
\multiput(22.5,17.5)(1,1){17}{\circle*{0.5}}
\put(37,32){\vector(1,1){3}}
\put(19,5){$\nu$}
\put(9,15){$|i\rangle$}
\put(65,15){\circle{5.5}}
\put(55,25){\circle{5.5}}
\put(53,27){\line(1,-1){4}}
\put(53,23){\line(1,1){4}}
\put(45,35){\vector(1,-1){7.5}}
\multiput(57.5,22.5)(1,-1){4}{\circle*{0.5}}
\put(59,21){\vector(1,-1){3}}
\multiput(67.5,17.5)(1,1){17}{\circle*{0.5}}
\put(82,32){\vector(1,1){3}}
\put(64,5){$\nu'$}
\put(54,15){$|f\rangle$}
\put(42,3){b}
\end{picture}
\end{center}
\bigskip

\begin{center}
\begin{picture}(85,35)
\put(20,15){\circle{5.5}}
\put(10,25){\circle{5.5}}
\put(8,27){\line(1,-1){4}}
\put(8,23){\line(1,1){4}}
\put(0,35){\vector(1,-1){7.5}}
\multiput(12.5,22.5)(1,-1){4}{\circle*{0.5}}
\put(14,21){\vector(1,-1){3}}
\multiput(22.5,17.5)(1,1){17}{\circle*{0.5}}
\put(37,32){\vector(1,1){3}}
\put(19,5){$\nu$}
\put(9,15){$|f\rangle$}
\put(65,15){\circle{5.5}}
\put(55,25){\circle{5.5}}
\put(53,27){\line(1,-1){4}}
\put(53,23){\line(1,1){4}}
\put(45,35){\vector(1,-1){7.5}}
\multiput(57.5,22.5)(1,-1){4}{\circle*{0.5}}
\put(59,21){\vector(1,-1){3}}
\multiput(67.5,17.5)(1,1){17}{\circle*{0.5}}
\put(82,32){\vector(1,1){3}}
\put(64,5){$\nu'$}
\put(54,15){$|f\rangle$}
\put(42,3){c}
\end{picture}
\end{center}
\bigskip

\centerline{Figure 1}

\end{document}